\documentclass[aps,prl,twocolumn,superscriptaddress]{revtex4-1}

\usepackage{color}
\usepackage{graphicx}
\usepackage{hyperref}

\newcommand{\jpsi}{\mathrm{J/}\psi}

\begin{document}

\title{
Collinearly improved kernel suppresses Coulomb tails in the impact-parameter dependent Balitsky-Kovchegov evolution.
}

\author{J. Cepila}
\affiliation{Faculty of Nuclear Sciences and Physical Engineering,
Czech Technical University in Prague, Czech Republic}
\author{J. G. Contreras}
\affiliation{Faculty of Nuclear Sciences and Physical Engineering,
Czech Technical University in Prague, Czech Republic}
\author{M. Matas}
\affiliation{Faculty of Nuclear Sciences and Physical Engineering,
Czech Technical University in Prague, Czech Republic}

\date{\today}

\begin{abstract}
We solved the impact-parameter dependent Balitsky-Kovchegov equation with the recently proposed collinearly improved kernel. We find that the solutions do not present the Coulomb tails that have affected previous studies. We also show that once choosing an adequate initial condition it is possible to obtain a reasonable description of HERA data on the structure function of the proton, as well as on the cross section for the exclusive production of a $\jpsi$ vector meson off proton targets. As a further application of the solutions, we computed the  impact-parameter dependent Weizs\"acker-Williams gluon distribution. 
\end{abstract}


\maketitle

\paragraph{\label{sec:Intro}Introduction.}
The high-energy, or equivalently small Bjorken-$x$, limit of perturbative Quantum Chromodynamics (pQCD) has received significant attention in recent years. From the experimental side, this has been driven by the precise measurements from HERA~\cite{Newman:2013ada}, the large kinematic reach of the LHC~\cite{N.Cartiglia:2015gve}, and the proposal of new electron-ion facilities~\cite{Accardi:2012qut,AbelleiraFernandez:2012cc}. In particular, the precise measurement of the  
$F_2(x,Q^2)$ structure function of the proton at HERA and its interpretation within pQCD~\cite{Aaron:2009aa,Abramowicz:2015mha} shows that the gluon distribution grows rapidly with decreasing $x$ for a fixed $Q^2$, where $x$ is fractional momentum of the struck parton and $Q^2$ is the negative squared four-momentum transferred between the lepton and the nucleon. This growth has to be tamed at some high energy in order to respect unitarity.

 In this limit, integro-differential equations are a powerful tool to compute and predict observables related to the dynamics of pQCD where the non-perturbative contributions are typically incorporated into an initial condition. In the seminal work~\cite{Gribov:1984tu}, it was shown that the inclusion of a non-linear term in these so-called evolution equations would limit the growth of the gluon distribution, a phenomenon known as saturation, see e.g.~\cite{Albacete:2014fwa} and references therein.
In this context, the Balitsky-Kovchegov (BK) equation~\cite{Balitsky:1995ub, JalilianMarian:1997gr} has been quite successful for phenomenological studies.
This equation was derived independently in the formalism of the operator product expansion in~\cite{Balitsky:1995ub} and within the dipole approach
in~\cite{Kovchegov:1999yj,Kovchegov:1999ua}. It can also be obtained within the Color Glass Condensate model as a limit of the so-called JIMWLK equations~\cite{JalilianMarian:1997gr,JalilianMarian:1997dw,Weigert:2000gi,Iancu:2000hn,Iancu:2001ad,Ferreiro:2001qy, Mantysaari:2018zdd}.

The BK equation describes the evolution with rapidity, $Y$, of the dipole-target scattering amplitude, $N(\vec{r},\vec{b};Y)$, where $\vec{r}$ is the transverse size of the dipole, $\vec{b}$ the impact parameter, and   $Y=\ln(x_0/x)$ with $x_0$ being the $x$ value at the start of the evolution. Solutions obtained under the assumption that there is no dependence on the impact parameter describe quite well the $F_2(x,Q^2)$ data~\cite{Albacete:2010sy}. This equation has also been solved including the impact parameter dependence~\cite{GolecBiernat:2003ym,Berger:2010sh}, where it was found out that the solutions acquired a so-called Coulomb tail, meaning that the contribution at large impact parameters grew too fast. This behavior was curbed by introducing an extra term to the kernel; furthermore, it was necessary to include an extra, so-called {\em soft}, contribution in order to describe $F_2(x,Q^2)$ data~\cite{Berger:2011ew}. With this approach it was also possible to describe the exclusive production of vector mesons in deeply inelastic scattering~\cite{Berger:2012wx}. These studies were based on a BK equation with a kernel including running coupling corrections~\cite{Balitsky:2006wa,Kovchegov:2006vj}. Recently, a new kernel including collinear corrections was proposed and shown to describe correctly HERA data on $F_2(x,Q^2)$ in an impact-parameter independent BK equation~\cite{Iancu:2015vea,Iancu:2015joa}.

In this work we study  the BK equation including the dependence on the impact parameter using the collinearly improved kernel. We find that  the Coulomb tails are strongly suppressed with respect to the running coupling case. Furthermore, we show that when using an appropriate initial condition a good description of experimental data is directly obtained; that is, without having to modify the kernel nor having to add extra soft contributions. 

The improved treatment of the impact parameter dependence provides a new tool for phenomenology. This tool is particularly important for the EIC facilities being currently under design and which have as one of their main goals a tomographic study of the structure of nucleons and nuclei~\cite{Accardi:2012qut,AbelleiraFernandez:2012cc}.

\paragraph{\label{sec:BKeq}The Balitsky-Kovchegov equation.}
We  assume a rotational symmetry of the target which implies that the scattering amplitude depends on the magnitude of the impact parameter, $b$, but not on its orientation. Furthermore, we assume the scattering amplitude to be independent of the angle between the vectors $\vec{r}$ and $\vec{b}$. In this case, the  BK  equation reads
\begin{eqnarray}
& & \frac{\partial N(r, b;Y)}{\partial Y} =  \int d\vec{r_{1}}K(r,r_{1},r_{2})\Big(N(r_{1}, b_1; Y) 
\nonumber \\
&&  + N(r_{2}, b_2;Y)- N(r, b;Y) - N(r_{1}, b_1, Y)N(r_{2}, b_2;Y)\Big), \nonumber\\
\label{fullbalitsky}
\end{eqnarray}
where $\vec{r_2}=\vec{r}-\vec{r_1}$, $|\vec{r}|\equiv r$ with similar definitions for  $r_1$ and $r_2$, while $b_1$ and $b_2$ are the magnitudes of the impact parameters of the respective dipoles. The collinearly improved kernel~\cite{Iancu:2015vea,Iancu:2015joa, Motyka:2009gi} is given by
\begin{equation}\label{collinearlyimproved}
K(r, r_1, r_2)  = \frac{\overline{\alpha}_s}{2\pi}\frac{r^{2}}{r_{1}^{2}r_{2}^{2}} \left[\frac{r^{2}}{\min(r_{1}^{2}, r_{2}^{2})}\right]^{\pm \overline{\alpha}_sA_1} \frac{J_1(2\sqrt{\overline{\alpha}_s \rho^2})}{\sqrt{\overline{\alpha}_s \rho}}.
\end{equation}
It constitutes of four factors. The factors $\overline{\alpha}_s/2\pi$ and $r^{2}/r_{1}^{2}r_{2}^{2}$ are present already at the LO, the factor in square brackets represents the contribution of single collinear logarithms and factor $J_1(2\sqrt{\overline{\alpha}_s \rho^2}) / \sqrt{\overline{\alpha}_s \rho}$ resums double collinear logarithms to all orders. Parameter $A_1= 11/12$ and the sign in the third factor is positive when $r^2 < \min(r_1^2, r_2^2)$ and negative  otherwise. 
$J_1$  is the Bessel function, $\rho\equiv\sqrt{L_{r_1r}L_{r_2r}}$ and $L_{r_ir} \equiv \ln(r_i^2/r^2)$.
For the running coupling,
$\overline{\alpha}_s \equiv \alpha_sN_c/\pi$ with $N_c$ the number of colors,
we use the smallest dipole prescription: $\alpha_s = \alpha_s (r_{\min})$,
where $r_{\min} = \min(r_1,r_2,r)$. This prescription has been used in previous studies, where it was compared to other prescriptions at a phenomenological level~\cite{Iancu:2015joa}; it has also been advocated to be the correct prescription for the BK equation at NLO~\cite{Balitsky:2008zza}.

To be consistent with the computations leading to the BK equation the form of the running coupling is given by 
\begin{equation}\label{alph}
\alpha_{s} (r) = \frac{4\pi}{\beta_{0,n_{f}}\ln\left(\frac{4C^{2}}{r^{2}\Lambda ^{2}_{n_{f}}}\right)},
\end{equation}

where $n_f$ denotes the number of flavors that are active at the scale $r$ and $\beta_{0, n_f}$ is the leading order coefficient of the QCD beta-series.  The value of $\Lambda^2_{n_f}$ depends on the number of active flavors and was computed in the same manner as in~\cite{Albacete:2010sy}. Two parameters control the infrared behavior of $\alpha_s$: $\alpha_{fr}$ and $C^2$. For very large dipoles the perturbative form of $\alpha_s$ given by Eq.~(\ref{alph}) is not anymore valid. Following the procedure used in previous studies~\cite{Albacete:2010sy} (see also discussion in Sec II.C of~\cite{Albacete:2009fh}) we freeze the value of $\alpha_s$ to $\alpha_{fr}$ = 1.0 for all dipole sizes that would produce a larger value of $\alpha_s$ when using Eq.~(\ref{alph}).  This is a purely phenomenological approach, which roughly describes the behavior found in more theoretical studies of $\alpha_s$ in the nonperturbative regime~\cite{Binosi:2016nme, Brodsky:2010ur}. Finally, the parameter $C^2$ also contributes to regulate the infrared behavior and takes into account the potential effect of the approximations made when computing the Fourier transform to coordinate space~\cite{Albacete:2007yr, Balitsky:2008zza}. 

\paragraph{\label{sec:Sol} Solving the BK equation.}
For the initial condition we use a combination of the GBW model~\cite{GolecBiernat:1998js} for the dependence on the dipole size $r$ and a Gaussian distribution for the impact parameter dependence. A similar approach has been considered in~\cite{McLerran:1997fk}. We use the following functional form
\begin{equation}\label{eq:initial_condition}
N(r, b,Y=0) =  1 - \exp\left(-\frac{1}{2}\frac{Q^2_s}{4}r^2 T(b_{q_1},b_{q_2})\right),
\end{equation}
where $b_{q_i}$ are the impact parameters of the quark and antiquark forming the dipole and
\begin{equation}\label{eq:impact_parameter_profile}
T(b_{q_1},b_{q_2})= \left[\exp\left(-\frac{b_{q_1}^2}{2B}\right) + \exp\left(-\frac{b_{q_2}^2}{2B}\right)\right].
\end{equation}
Both $Q^2_s$ and $B$ are parameters to be adjusted. These parameters have a clear interpretation: the scale at which nonlinear effects become important, known as the saturation scale, is given by $Q^2_s$; while $B$ is related to the effective radius of the Gaussian distribution in impact parameter space that represents the target profile by $2B=\langle b^2\rangle$. $T(b_{q_1},b_{q_2})$ suppresses contributions from dipoles that are large with respect to the size of the target. Such suppression of large dipole sizes, which makes sense from the phenomenological point of view, has also been used in previous approaches~\cite{Berger:2011ew} in order to describe the data. 

Parameter $B$ was chosen to obtain a reasonable description of  the cross section for $\jpsi$ photoproduction off protons as a function of $|t|$ ($-t$ is the square of the momentum transferred at the proton vertex) at a fixed center-of-mass energy of the photon--proton system ($W=100$ GeV), while $Q^2_s$ was simultaneously chosen to describe $F_2(x,Q^2)$ data at $x_0=0.008$ and $Q^2\in(3.5,27)$ GeV$^2$. That is, the fixing of $Q^2_s$ does not involve an evolution in $Y$, while that of $B$ requires evolving the dipole scattering amplitude to $x\approx0.001$. (This value is obtained from $x=(M_{\jpsi}/W)^2$ where $M_{\jpsi}$ is the mass of the $\jpsi$.) These two conditions uniquely fix the value of these two parameters, since the structure function is sensitive to an overall integral of the scattering amplitude and vector meson production is sensitive to the $b$-dependence of it. The values we use in the following are $Q^2_s$ = 0.49 GeV$^2$ and  $B$ = 3.22 GeV$^{-2}$. The value of $C^2$ used in the computation of $\alpha_{s} (r)$ was chosen to regulate the evolution speed of the dipole scattering amplitude and set to $C=9$.

The BK equation is solved numerically using the Runge-Kutta method of order four with the algorithm  described in~\cite{Cepila:2015qea,Matas:2016bwc}, extended to include the $b$-dependence. The grids in $\log_{10}(r)$ and $\log_{10}(b)$ are of the same size and cover the range from $10^{-7}$ to $10^2$~1/GeV for both $r$ and $b$. A linear interpolation in  $\log_{10}(r)$ and $\log_{10}(b)$ is used to find the value of the dipole scattering amplitude outside the points in the grids. The step in rapidity was 0.01. The integrals are performed with the Simpson method.  

\begin{figure}[t!]
  \centering
   \includegraphics[width=\linewidth]{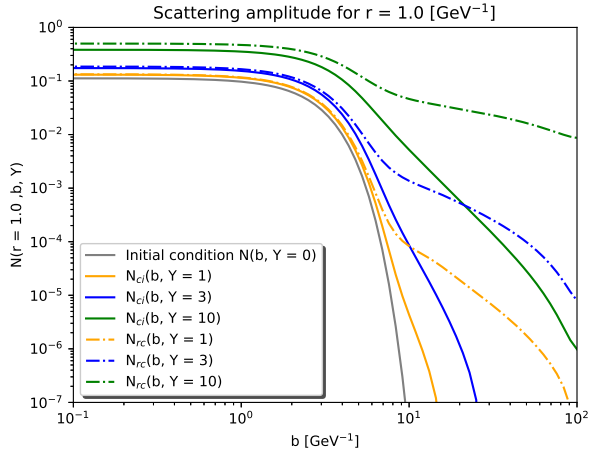}
\caption{(Color online) Dependence of the dipole scattering amplitude with respect to the impact parameter at different rapidities for a dipole of size $r=1$ GeV$^{-1}$. The dashed-dotted lines represent solutions obtained with the running-coupling kernel ($N_{rc}$), while solid lines represent solutions with the  collinearly improved kernel ($N_{ci}$).}
\label{fig:bdep_vs_b}
\end{figure}

Using the procedure just described we obtained the solutions presented in Fig.~\ref{fig:bdep_vs_b}, which shows the impact-parameter dependence of the dipole scattering amplitude for a dipole of size $r=1$ GeV$^{-1}$ at different rapidities for two computations: using the collinearly improved or the running-coupling kernel. In both cases we use the same initial condition. 

We show results for rapidities which are relevant for phenomenology at current and planned facilities, but have checked that such a behavior is still present even at Y=10, which is beyond the reach of foreseeable accelerators. The evaluation of $\alpha_s(r)$ for the running coupling case is done as in~\cite{Albacete:2010sy}. Figure~\ref{fig:bdep_vs_b} shows that the Coulomb tails are strongly suppressed when using the collinearly improved kernel. A similar pattern is observed for all dipole sizes. 
The suppression of the amplitude at large values of $b$ observed when using the collinearly improved kernel instead of the running coupling kernel is due to two reasons: ($i$) the different treatment of the $r^{2}/r_{1}^{2}r_{2}^{2}$ factor, which in the running coupling kernel appears accompanied by other additive terms, and ($ii$) the new corrections introduced in the collinearly improved kernel. When comparing the original LO with the collinearly improved kernel, there are three factors contributing to the suppression: the use of a running coupling constant instead of a fixed $\alpha_s$, the contribution of single collinear logarithms, and the resummation of double collinear logarithms. This last term is numerically the most important.
A detailed discussion of the properties of the solutions found with our approach is outside the scope of this work and will be presented elsewhere~\cite{CCM}.

\paragraph{\label{sec:App} Applications.}
As a first use of the solutions to the $b$-dependent BK equation we compute the $F_2(x,Q^2)$ structure function and compare the result with HERA data. In the dipole model  the structure function is related to the dipole scattering amplitude by

 \begin{eqnarray}\label{F22}
  F_{2}(x, Q^{2}) &=& \frac{Q^{2}}{4\pi^{2}\alpha_{\rm em}}\sum_{f}  \int {\rm d}\vec{r}\;{\rm d}\vec{b}\;{\rm d}z\nonumber \\
  && \mid\Psi_{T,L}^{f}( z, \vec{r}\,)\mid^{2}  \frac{{\rm d}\sigma^{q\bar{q}}(\vec{r}, x_f)}{{\rm d}\vec{b}},
\end{eqnarray}
where $\alpha_{\rm em}$ is the electromagnetic coupling constant,  $\Psi_{T,L}^{f}( z, \vec{r}\,)$ is the convolution of the wave functions for a photon to split into a quark-antiquark dipole of flavor $f$ and for the dipole to return to the photon state --- see e.g.~\cite{Kowalski:2006hc} for a detailed discussion ---, $z$ is the fraction of the dipole energy carried by the quark, and the cross section is related to the dipole scattering amplitude by
 \begin{equation}
 \frac{{\rm d}\sigma^{q\bar{q}}(\vec{r}, x)}{{\rm d}\vec{b}} = 2N(\vec{r}, \vec{b}, x).
 \label{eq:qq_xs}
 \end{equation}
 As  it is customary, we use  $x_f = x ( 1 + (4m^2_f)/Q^2)$ with $m_f$ an effective quark mass set to 100~MeV/$c^2$ for light quarks. The description of data shown below does not depend strongly on the value of $m_f$ and remains the same if a value of 10 MeV/$c^2$ is used. Similar observations were made in~\cite{Iancu:2015joa}. In the future, it would be interesting to match this prescription with a more formal description of dressed quarks as e.g. in~\cite{Bhagwat:2003vw}. Mass of the charm quark was fixed to 1.3~GeV/$c^2$; these values are the same as used in~\cite{Iancu:2015joa}.

\begin{figure}[t!]
  \centering
   \includegraphics[width=\linewidth]{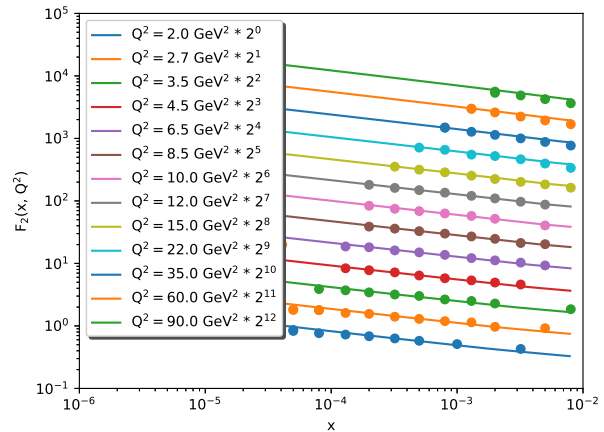}
\caption{(Color online) Comparison of the structure function data from HERA~\cite{Aaron:2009aa} with the computation based on solutions to the collinearly improved $b$-dependent BK equation.}
\label{fig:bdepbF2}
\end{figure}

Figure~\ref{fig:bdepbF2} shows the comparison of the computation with the measured data~\cite{Aaron:2009aa} for several different values of $Q^2$ as a function of $x$. The average percentile difference between data and theory is 3.7\% for data with $Q^2\in[3.5,35]$ GeV$^2$. We would like to emphasize that this level of agreement was obtained without the need to include ad hoc corrections to the kernel and without the addition of soft contributions.  

As a further application we computed the $|t|$ dependence of cross section for the exclusive photoproduction  of $\jpsi$ vector mesons off protons at fixed values of $W$. The amplitude for this process is given by (see e.g.~\cite{Kowalski:2006hc})
\vfill\break
\begin{eqnarray}
A(x,Q^2,\vec{\Delta})_{T,L} &=& i\int {\rm d}\vec{r}\int^1_0\frac{{\rm d}z}{4\pi}
(\Psi^*\Psi_{\jpsi})_{T,L} \nonumber \\
& &\int {\rm d}\vec{b}\;
e^{-i(\vec{b}-(1-z)\vec{r})\cdot\vec{\Delta}}\frac{{\rm d}\sigma^{q\bar{q}}}{{\rm d}\vec{b}},
\label{eq:Amplitude}
\end{eqnarray}
where $-t\equiv\vec{\Delta}^2$, $T$ and $L$ represent transverse and longitudinal photons, respectively, and $\Psi_{\jpsi}$ is the wave function of the transition from the dipole into a $\jpsi$ vector meson. We use the boosted Gaussian wave functions~\cite{Nemchik:1994fp, Nemchik:1996cw} with parameters as determined in~\cite{Kowalski:2006hc}. 

The $|t|$-differential cross section is given by the square of the amplitude divided by 16$\pi$. The contributions from the longitudinal and transverse photons are added. As it is customary (see discussion in Sec. 3 of~\cite{Kowalski:2006hc}), we correct the cross section for two effects: ($i$) to take into account the contribution of the real part of the dipole scattering amplitude that was not considered when deriving the form of the amplitude in Eq.~(\ref{eq:Amplitude}), and ($ii$) the fact that in a two-gluon exchange the gluons have different momentum, which is known as the skewedness correction~\cite{Shuvaev:1999ce}. The correction has been computed using the derivative of the amplitude as in~\cite{Kowalski:2006hc}. The correction in this context has to be understood as a phenomenological ingredient that contributes up to a value of 30~\% to the total cross section.

 \begin{figure}[t!]
  \centering
   \includegraphics[width=\linewidth]{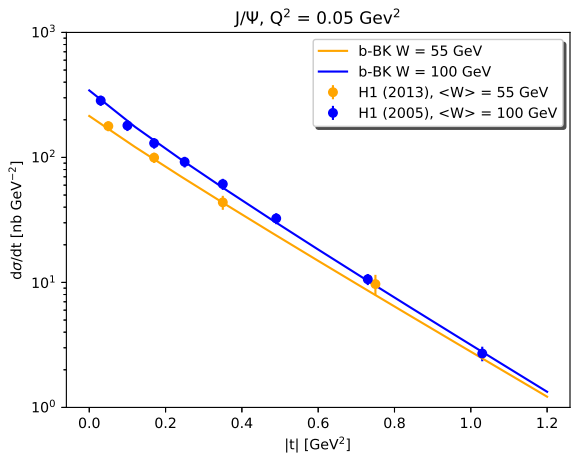}
\caption{(Color online) Comparison of the computation for the $|t|$ dependence of the cross section for the exclusive photoproduction of $\jpsi$ vector mesons off protons with data from the H1 Collaboration at HERA at $\langle W\rangle$ = 55\,GeV~\cite{Alexa:2013xxa} and  $\langle W\rangle$ = 100\,GeV~\cite{Aktas:2005xu}. }
\label{fig:jpsi-t_cs}
\end{figure}

The comparison of the computation with data from the H1 Collaboration~\cite{Aktas:2005xu,Alexa:2013xxa} is shown in Fig.~\ref{fig:jpsi-t_cs}. Note that the data at $\langle W \rangle = 100$ GeV were used to set the value of the parameter $B$, but the computation for $W=50$ GeV is a prediction. The agreement is at the level of 10\%.

As a final application of the dipole scattering amplitude solutions  to the $b$-dependent BK equation with the collinearly improved kernel we turn to  TMD (transverse momentum dependent) distributions. The measurement of these distributions is one of the goals of future facilities which are being currently designed~\cite{Accardi:2012qut,AbelleiraFernandez:2012cc}. There are also recent ideas on how to access this kind of distributions, and how to apply them to phenomenology, using LHC data, see e.g.~\cite{vanHameren:2016ftb,Hagiwara:2017fye,Albacete:2018ruq}. Here, as an example of the potential of the solutions we found,
we compute the impact-parameter dependent Weizs\"acker-Williams  gluon distribution $G^{(1)}$.

This gluon distribution can be interpreted as the number density of gluons at certain $x$ and with a given transverse momentum, $k_t$,  at a distance  $b$ from the center of the proton. Its relation to the dipole scattering amplitude as given in~\cite{vanHameren:2016ftb} is (see e.g.~\cite{Marquet:2016cgx})
\begin{eqnarray}
\alpha_s xG^{(1)}(x,k_t, b) &=& \frac{N_c}{4\pi^4}\int \frac{{\rm d}\vec{r}}{r^2}\
e^{-i\vec{k_t}\cdot\vec{r}}\nonumber \\
& & \left\{1 - [1-N(x,r, b)]^2\right\}.
\label{eq:WW_AdjointDipole}
\end{eqnarray}

\begin{figure}[t!]
  \centering
  \includegraphics[width=\linewidth]{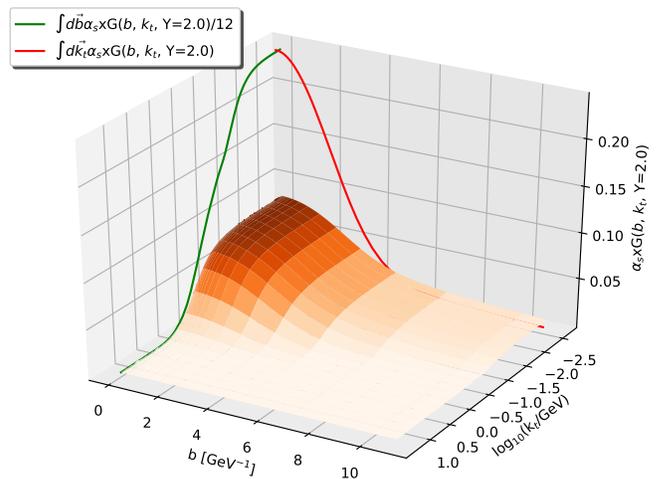}
\caption{(Color online) The impact-parameter dependent Weizs\"acker-Williams gluon distribution computed from the solution to the BK equation with the collinearly improved kernel. The red and green lines represent the integral of this distribution over the transverse momentum $\vec{k_t}$ or over the impact parameter $\vec{b}$, respectively. }
\label{WW0}
\end{figure}

Figure~\ref{WW0} shows the impact-parameter dependent Weizs\"acker-Williams gluon distribution computed with the dipole scattering amplitude obtained as a solution to the $b$-dependent BK equation with the collinarly improved kernel. The distribution is shown at a rapidity $Y=2$. The figure also shows the integrals of this distribution over $\vec{k_t}$ and over $\vec{b}$. Integrals of this distribution feature reasonable size in impact parameter and fast-falling dependence on $k_t$ (with an asymptotic behavior close to a power-like fall off with a power of -2, which was also reported in~\cite{vanHameren:2016ftb}), suggesting that these distributions are ready to be used for phenomenological studies.

\paragraph{\label{sec:Summary}  Summary and outlook.}
In this work we obtained the dipole scattering amplitude as a solution to the impact-parameter dependent Balitsky-Kovchegov equation using the collinearly improved kernel. We find that the Coulomb tails that have affected previous studies are strongly suppressed when using this kernel. Furthermore, we show that choosing specific initial conditions we obtain a good description of data on the  $F_2(x,Q^2)$ structure function of the proton and on the cross section for the $|t|$ dependence of exclusive photoproduction of $\jpsi$ vector mesons off protons. The agreement with data is obtained without the need of adding any extra term to the kernel and without any soft contribution. The success of these dipole scattering amplitudes in the description of data makes them valuable tools for phenomenological studies either using existing HERA and LHC data or to predict observables  for future colliders. In this context we presented first results on the impact-parameter dependent Weizs\"acker-Williams gluon distribution. 

As a last remark, we would like to point out that there have been important advances in the computation of the BK equation at the next order in perturbation theory. The new equation, presented in~\cite{Balitsky:2008zza}, has been solved in~\cite{Lappi:2016fmu} using the collinearly improved kernel, but without considering the impact parameter dependence. Furthermore, the tools to be able to use this equation for phenomenological applications are being developed, see e.g.~\cite{Beuf:2014uia,Beuf:2017bpd,Ducloue:2017ftk,Hanninen:2017ddy}. Our results indicate that solutions of the NLO-BK equation including the collinearly improved kernel and considering the impact-parameter dependence may be useful to understand better the properties of pQCD in the high-energy limit.

The dipole scattering amplitudes computed in this work  are publicly available in the website 

\url{https://hep.fjfi.cvut.cz/}.

\paragraph*{Acknowledgments.}
We would like to thank Dagmar Bendov\'a, Heikki M\"{a}ntysaari and Cyrille Marquet for fruitful discussions.
Our work has been partially supported by grant 17-04505S of the Czech Science Foundation, GA\v{C}R and the COST Action CA15213 THOR.  Computational resources were provided by the CESNET LM2015042 grant and the CERIT Scientific Cloud LM2015085, provided under the program "Projects of Large Research, Development, and Innovations Infrastructures". 
\bibliography{Biblio}

\end{document}